\documentstyle[12pt,fleqn]{article}
\renewcommand{\baselinestretch}{1.2}

\newcommand{\bea}{\begin{eqnarray}}
\newcommand{\beq}{\begin{equation}}
\newcommand{\eea}{\end{eqnarray}}
\newcommand{\eeq}{\end{equation}}
\newcommand{\nnu}{\nonumber}
\newcommand{\di}{\mbox{d}}
\newcommand{\g}{\tilde{\phi}}
\newcommand{\h}{\hat}
\newcommand{\n}{\nabla}
\newcommand{\al}{\alpha}
\newcommand{\be}{\beta}

\newcommand{\ga}{\gamma}
\newcommand{\half}{\frac{1}{2}}

\newcommand{\spav}[1]{\parbox{1mm}{\vspace*{#1}}}

\begin{document}

\begin{titlepage}
\begin{flushright}
SISSA/ISAS 172--91--EP\\
(revised version) April 1992
\end{flushright}
\spav{1cm}\\
\begin{center}
{\LARGE\bf \sc Strong Anti-Gravity\\}
{\bf \sc Life in the Shock Wave\\}
\spav{1.5cm}\\
{\large  Marco Fabbrichesi and Kaj Roland}
\spav{2cm}\\
{\em International School for Advanced Studies
(SISSA/ISAS)}\\
and \\
{\em INFN, Sezione di Trieste}
\spav{.5cm}\\
{\em Strada Costiera 11, I-34014 Trieste, Italy.} \\
\spav{2cm}\\
{\sc Abstract}
\end{center}

Strong anti-gravity is the vanishing to all orders in
Newton's constant of the net force
 between two massive particles at rest.
 We study this
phenomenon and show  that it occurs in any effective
theory  of gravity
which is obtained from a higher-dimensional model
 by compactification on a manifold with flat directions. We find the exact
 solution of
   the Einstein equations  in the presence of a point-like
   source of strong
 anti-gravity by dimensional reduction of what is a
   shock-wave solution in the higher-dimensional model.
\vfill
\end{titlepage}

\newpage
\setcounter{footnote}{0}
\setcounter{page}{1}

\section{Introduction.}

\hspace*{2em}  A distinctive feature
of gravity is that it is always attractive and therefore
impossible to shield. This property is currently understood in terms of
the spin of the graviton, which is two and therefore
 an even number---the rule being
 that the exchange of particles of even spin gives rise
to forces that are always attractive, whereas
the exchange of particles
of odd spin gives forces  which are
attractive for charges of opposite sign and repulsive
for charges of the same sign.

By contrast, the term {\em anti-gravity} has come to represent all those
physical phenomena in which the usual
 gravitational potential---the
static limit of which is Newton's
inverse-square law---is modified to accommodate
repulsive gravitational forces.
A simple example of such a modification is a
theory in which anti-matter   gravitationally  repels ordinary matter the
same way as  electric charges of the same sign do.

This is clearly a
fascinating subject with many implications:
from the possible
 check of anti-gravity against experiments to the several theoretical
 issues that are involved, the principle of equivalence and energy
 conservation among others (see~\cite{Nieto} for a recent and
  comprehensive review of the subject).

However, in the present paper the term anti-gravity
is used in the more  circumscribed sense proposed by
J. Scherk~\cite{Scherk}: it
stands for a theory in which the
gravitational  potential between two masses  at rest
vanishes in the Newtonian approximation
because  in addition to the attractive exchange of
spin-2 gravitons there  also exists a  repulsive
contribution coming from  the exchange of
spin-1 particles, the {\em graviphotons}.
  Matter and anti-matter  behave in
the ordinary manner under the exchange of the gravitons but,  whenever
appropriately
charged, also
couple to the graviphotons (and, possibly, to spin-0
{\em graviscalars} as
well). The respective couplings are arranged to give a vanishing
net force.

 Such a
fine tuning of the coupling strengths, as contrived as it may seem at
first,  was found to take place
 inside $N=2$, $D=4$ supergravity
by Zachos~\cite{Zachos}.  Other examples
were soon pointed out within the supergravity
family by  Scherk~\cite{Scherk}, and anti-gravity was accordingly
promoted  from  being a mere
curiosity to a potentially interesting
phenomenon.

In this paper we  reconsider  Scherk's
anti-gravity.  We
  show, first of
all,  that it
actually comes in two varieties: {\em weak} anti-gravity, like
in the $N=2$ model, in which the
vanishing of the static potential does not persist in the non-linear
theory, and {\em strong} anti-gravity, like in the $N=8$ model,
in which it does. Whereas
the  weak kind
seems to be of a more accidental nature, the strong one is a
widespread
phenomenon. It takes place (at sufficiently small distances) whenever:
\begin{itemize}
  \item
   the $D$
dimensional theory is obtained after compactification of some higher
dimensional model; and
\item
the states for which it occurs are mass\-less and neu\-tral
in the higher-di\-men\-sio\-nal theory but gain mass and charge in the process
of compactification.
\end{itemize}
{}From supergravity to string
theory,  most
modern theories of quantum gravity belong to
 such a class of models---a fact that
speaks for the relevance of anti-gravity.

Supersymmetry,
 as we shall discuss, is not among the requirements
 necessary for anti-gravity to take place, even though (when
 present) it allows an elegant characterization of the phenomenon. It
 also played a historical role inasmuch as all models of anti-gravity were
 first found inside supersymmetrical theories.

The very general behavior outlined above, as well as the extent to which it
applies,
 cry out for a simple physical explanation.
This can be found in the higher dimensional theory---as it was realized
already by Scherk himself~\cite{Scherk}: gravitons,
graviphotons and graviscalars are just different components of the
higher-dimensional metric tensor and the anti-gravity phenomenon follows
directly from the well-known laws of gravitational interactions of
massless particles. In particular, the exact solution of an
anti-gravitating source is just a  generalized
shock wave moving in one of the
compact directions. The $D$ dimensional world lives
so-to-speak inside such a shock wave.
 By turning the argument around, one can say that
strong anti-gravity is only a complicated way to describe in $D$
dimensions what is a very simple metric in $D+E$.
  Nonetheless,
in the process
something remarkable has occurred: we have found an exact solution of
Einstein's equations in $D$ dimensions for  a theory
 in which graviphotons and graviscalars are present
together with the gravitons. In four dimensions it is a solution of the
complete
field equations
of $N=8$  (or $N=4$) supergravity.

This anti-gravity solution---which we arrived at through the shock-wave
analogy outlined above---corresponds in $D$ dimensions to
a static and spherically symmetric
solution of gravity coupled    to
Maxwell and  scalar fields for  an  extremal
value of the respective charges. These charged black hole solutions
have
been discussed
 in refs.~\cite{Gibbons,Chodos,Horowitz} and
 in~\cite{Gibbons,Pollard} with
 reference to anti-gravity. Ref.~\cite{Pollard} discusses
 the Newtonian limit
of the solutions of~\cite{Chodos} to check for the
presence of weak anti-gravity.

In this paper we study the motion of a charged test particle in the
anti-gravity background and find a complete and very simple solution to
the equations of motion. In particular, we prove that the static
potential vanishes not just in the Newtonian
limit  but at arbitrarily short
distances from the source, that is, compactification of pure gravity does
indeed lead to a theory of strong anti-gravity.

We proceed to establish the most general conditions under which strong
anti-gravity persists at long distances.
In most cases the graviphotons and graviscalars
will be massive below the scale of compactification. While
the existence
of a Killing vector on the compact manifold ensures the
presence of a massless graviphoton, the appropriate graviscalar
will be massive unless
 the
manifold has
 a flat direction.  More specifically, anti-gravity is a
 long-distance phenomenon in $D$ space time dimensions if
 \begin{itemize}
 \item
 the extra $E$ dimensions are compactified
 on a manifold that is Ricci flat
 and of holonomy SO($E-1$); only compact manifolds with a flat direction
satisfy
 these requirements.
 \end{itemize}

 The physical viability of anti-gravity
 can be discussed in the light of
 our results.
  Calabi-Yau
 manifolds, the holonomy of which is SU(3),
 do give a mass to the
   graviphotons~\cite{GSW}
and, therefore, in these models anti-gravity is
   limited to energies larger than the compactification scale.  Below
   that scale, graviphotons and graviscalars give rise to Yukawa-like
   short-range forces and effectively decouple from the gravitational
   interactions.

 The content of the paper is as follows. In section 2 we
 review the two models first proposed by Scherk, that is,
 $N=2$ and $N=8$
supergravity and add to them a model inspired by the superstring.
Section 3 contains a study of anti-gravity in the more general setting
outlined above: the
Einstein-Hilbert action is compactified on tori and the matter fields are
taken to be massless multiplets with momenta in one
or more of the compact
directions. We find anti-gravity by considering the Newtonian potential for
the interaction of these states. In section 4
we write the exact solution of such an anti-graviting source and discuss the
corresponding motion of a test particle. In section 5  we consider
the
most general compactification scheme that allows long-range
anti-gravity.
Finally, section 6 contains our conclusions. We have tried to include all
the material necessary to make the paper as self-contained as possible.

\section{Three examples of anti-gravity.}
\setcounter{equation}{0}
\hspace*{2em} In this section we  briefly review the two models of
anti-gravity originally discussed by Scherk~\cite{Scherk}
 and add one of our own:
 type-II
superstring theory with toroidal compactification.

Of the terms entering the static
potential, only  those
linear in Newton's constant $G_N$
are here taken into account. A discussion of the same models in the fully
non-linear theory is postponed to section 4.

\subsection{The N=2 model.}
\hspace*{2em} This model was first discussed by Zachos~\cite{Zachos}.
The graviton multiplet in $D=4$, $N=2$ supergravity
contains, besides the graviton
and the two gravitinos, a single vector field. Within the framework of field
theory the only way to introduce massive matter is by means of a multiplet
with central charge
\beq
Z = 2m  \label{2.1}\, ,
\eeq
where $m$ is the mass of the multiplet. This  is
precisely the value of the central
charge which
reduces the dimension of the representation to that of the massless
representation~\cite{3b} and thus
ensures that only matter fields of spin 0 and spin 1/2
enter. The vector field gauges the central charge. The corresponding value
of the charge is
\beq
q =  \kappa m / \sqrt{2} \label{2.2}\, ,
\eeq
where $\kappa^2 = 8\pi G_N$. The value (\ref{2.2}) for the charge
is exactly the one needed in the static limit
to obtain a cancellation between
gravitational  and  ``electric'' forces. The static Newtonian
potential is
\beq
 V(r) = - \frac{\kappa^2}{8 \pi} \frac{m_1 m_2}{r} ( 1 - \epsilon_1
\epsilon_2 )  \label{2.4a}\, ,
\eeq
where $\epsilon = +1$ for particles and $-1$ for antiparticles. In
eq.~(\ref{2.4a}) the cancellation is  between the attractive spin-2
graviton and the the repulsive spin-1 graviphoton. Because
no other particles partake in the interaction, $N=2$
is the simplest instance of anti-gravity one can think of.

\subsection{The N=8 model.}
\hspace*{2em} The second example of  a theory with
anti-gravity  is $N=8$,  $D=4$ supergravity, a model extensively
studied by Scherk~\cite{Scherk}.

 The  $N=8$,  $D=4$
 supergravity theory~\cite{3c} can be obtained by dimensional reduction
from $N=1$ supergravity in eleven dimensions~\cite{3d}.
All massless particles are unified in the graviton multiplet, which
contains, besides the graviton and the 8 gravitinos,  28 spin 1,
56 spin 1/2 and 70 spin 0 particles. It is not possible (within the
framework of supersymmetric quantum field theory) to couple this theory
to any kind of matter, because any massive multiplet contains (at least)
particles with spins  up to two
and it is not known how to write consistent field
theories for massive particles with spin greater than one.

Scherk~\cite{Scherk} obtains massive matter from the graviton multiplet itself
by means of a generalized dimensional reduction~\cite{5a,5b}
that breaks
the $N=8$ supersymmetry.
The theory is first reduced to unbroken $N=8$ supergravity in five
dimensions by ordinary dimensional reduction.
Then---in going
from five to four dimensions---the global SO(6) invariance
present in the spectrum of that theory (which describes global rotations of
the six coordinates already compactified) can be used to introduce a
dependence on the  coordinate $x^5$:
\beq
\phi(x^{\mu}, x^5) = \exp( i M x^5 ) \phi (x^{\mu})  \label{2.3}\, ,
\eeq
where $M$ is an element of the SO(6) Lie algebra containing three arbitrary
parameters with the dimension of a mass. The field $\phi$ becomes
multivalued in going around the fifth direction but
the ambiguity only amounts to a symmetry
transformation.
This way, any field in $D=5$ that transforms non-trivially under  SO(6)
acquires a mass in $D=4$
\cite{5b}\footnote{The global SO(6) symmetry can actually
be extended to a global E(6) with maximal compact subgroup Sp(8) of rank
four. This way, a fourth mass parameter can be introduced~\cite{CSS}.}.

Among the fields that remain massless is the
five-dimensional graviton field.
In four dimensions this decomposes into a graviton, a spin-1  graviphoton
and a spin-0  graviscalar. The static potential  between any
massive states of like charges vanishes
by virtue of the repulsion of the
graviphoton balancing the total attraction due to
the graviton and graviscalar:
\beq
V(r) = - \frac{\kappa^2}{8 \pi} \frac{m_1 m_2}{r} ( 1 - 4 \epsilon_1
\epsilon_2 + 3)  \label{2.4}\, ,
\eeq
 where the terms in
the bracket arise from spin-2, spin-1 and spin-0 exchange, respectively.

\subsection{The type-II superstring.}
\hspace*{2em} Modern
string theory was still to come when anti-gravity first appeared.
It is however straightforward to
add  to the
preceding two examples a model based on the superstring. This boils down to
another way of introducing massive matter in the
$N=8$,  $D=4$ model.

The $N=8$ model can be obtained from an
$N_+ = N_- = 1$ supergravity in $D=10$
dimensions. This theory in turn can be considered the low-energy limit of
a type-II superstring~\cite{GSW}.
Now, states with non-vanishing momentum
in the compact direction will be
seen as massive states in four dimensions, as it is always the case in
Kaluza-Klein theories.

At this point,  it is important to
 distinguish between states which are massive already in
ten dimensions (i.e. massive excitations of the string) and states that are
massless in ten dimensions and have only  a mass in four
dimensions by virtue of a nonzero compact momentum.

The difference between these two kinds of states
shows up in  their
 static interactions, which are easily obtained from the
 four-point  Veneziano amplitudes in the limit $t \rightarrow 0$
(where $t$  and $s$ below are the usual Mandelstam variable).
For instance, the elastic scattering of two massive string states
\beq
|\Psi_i \rangle = B^i_{\mu \nu \rho ; \bar{\mu} \bar{\nu} \bar{\rho}}
\left( \psi^{\mu}_{-1/2} \psi^{\nu}_{-1/2} \psi^{\rho}_{-1/2}\right)
\: \left( \bar{\psi}^{\bar{\mu}}_{-1/2} \bar{\psi}^{\bar{\nu}}_{-1/2}
\bar{\psi}^{\bar{\rho}}_{-1/2} \right)  |k;0\rangle \, ;
 \:\:\:\: i=1,2  \label{2.6}
\eeq
with mass $\alpha' m_{10}^2 = 4$ gives the amplitude
\beq
A(1234)  \sim  - \kappa^2 \frac{(s - 2 m_{10}^{2})^{2}}{t}\,
B_1 \cdot B_2 \, B_3 \cdot B_4  \label{2.7}\, ,
\eeq
where all the momenta are taken to be four-dimensional.
In the static limit
the amplitude~(\ref{2.7})
is nonzero and these states do not  produce any  anti-gravity.

If we consider instead the elastic scattering of two massless states
\beq
|\Psi_i \rangle = e^i_{\mu \bar{\mu}} \psi^{\mu}_{-1/2}
\bar{\psi}^{\bar{\mu}}_{-1/2} |k;0 \rangle  \, ; \:\:\: i=1,2  \label{2.8}
\eeq
with a conserved compact momentum $p^{i}_5$ in the fifth direction,
we obtain in four dimensions and in the static limit
\beq
A(1234) \sim  -\frac{4\kappa^2}{t} m^{2}_{1}m^{2}_{2}
(1 - \epsilon_1 \epsilon_2 )^{2} \, e_1 \cdot e_2 \: e_3 \cdot e_4
\, ,  \label{2.9}
\eeq
which is equivalent to (\ref{2.4}).
The states~(\ref{2.8}) therefore do give rise to
anti-gravity.

\section{The essence of anti-gravity.}
\setcounter{equation}{0}
\hspace*{2em} The three theories considered in the previous section are
very different,  differing in the content of their massless multiplet as
well as in the way the massive states are introduced. Yet all
three of them contain anti-gravity.  This fact indicates that some
general principle is at work. In this section we  prove that
anti-gravity is a general feature of any theory in $D$ dimensions which is
obtained by toroidal compactification from a theory in $D+E$ dimensions
with the  following two properties:
\begin{itemize}
\item
It contains gravity, described at long distances by the Einstein-Hilbert
action;
\item
the low-dimensional massive states descend from
 massless states which in the Born approximation couple only
gravitationally to each other.
\end{itemize}

\subsection{Compactification of Einstein theory.}
\hspace*{2em} To prove the above statement we first consider the
dimensional reduction of the Einstein-Hilbert action in $D+E$ dimensions:
\beq
\hat{S} = \frac{1}{2\kappa^2} \int \frac{\di ^{D+E} \hat{x}}{\rho (E)}
\sqrt{-\hat{g}}\: \hat{R} \label{3.1} \, .
\eeq
Before discussing the action~(\ref{3.1}) it is necessary to introduce
some notations.
Here, and in the following sections,
 $(D+E)$-dimensional objects carry a hat. The
$(D+E)$-dimensional vector $\hat{x}^{\hat{\mu}}$ is decomposed as
\beq
\hat{x}^{\hat{\mu}} = (x^{\mu} ; y^{\al} )
\eeq
with $\mu = 0, \ldots , D-1$ and $\al = 1, \ldots , E$. We shall refer to
$D$-dimensional objects as {\em external} and  $E$-dimensional as {\em
internal}.
External indices are denoted  by letters ($\mu$, $\nu$, $\rho$,
...) from the middle of
the Greek alphabet; internal ones by letters ($\al$, $\be$, $\gamma$, ...)
from the beginning of it.
$\rho (E)$ is the {\em coordinate} volume of the toroidal manifold.
If the coordinate $y^{\al}$ takes values in the interval $[0; L^{\al}]$
(which is taken to be fixed under reparametrization) then
\beq
\rho (E) = \prod_{\al = 1}^{E} L^{\al} \, .
\eeq
This volume factor is inserted into (\ref{3.1}) to ensure that
$\kappa^{2}$ is the gravitational constant in $D$ rather than
in $D+E$ dimensions. Our conventions for the metric and curvature are
those of~\cite{MTW}.

The $y$-dependence of the $(D+E)$-dimensional metric
$\h{g}_{\h{\mu}\h{\nu}}$ can be
expanded on Fourier modes. The zero mode (corresponding to no dependence
on $y$) gives  massless states in $D$ dimensions.
As in ref.~\cite{5b}, we introduce the following decomposition for the
metric tensor:
\begin{displaymath}
\h{g}_{\h{\mu}\h{\nu}} = \left(
\begin{array}{ll}
\delta^{\ga}g_{\mu\nu} + 2\kappa^{2}A^{\al}_{\mu}A^{\be}_{\nu}\phi_{\al\be}
& -\sqrt{2} \kappa A_{\mu}^{\be}\phi_{\al\be}   \\
-\sqrt{2} \kappa A_{\nu}^{\al}\phi_{\al\be} & \phi_{\al\be}
\end{array}
\right)
\end{displaymath}
\beq
\h{g}^{\h{\mu}\h{\nu}} = \left(
\begin{array}{ll}
\delta^{-\ga}g^{\mu\nu}
& \sqrt{2} \kappa \delta^{-\ga}A^{\mu\al}\\
\sqrt{2}\kappa  \delta^{-\ga} A^{\be\nu} & \phi^{\al\be}
+ 2\kappa^{2}\delta^{-\ga} A^{\al\mu}A^{\be}_{\mu}
\end{array}\label{3.3}
\right) \, .
\eeq
Apart from the $D$-dimensional metric
$g^{\mu\nu}$, the decomposition  (\ref{3.3}) introduces a
total of $E$ graviphotons $A^{\al}_{\mu}$ and $\frac{1}{2}E(E+1)$
graviscalars $\phi_{\al\be}$---the latter being  the components
of the compact space metric (whose inverse is $\phi^{\al\be}$).
In eqs.~(\ref{3.3})
 all external indices are raised and lowered by means of
$g^{\mu\nu}$,  the internal ones by $\phi_{\al\be}$. Furthermore,
\beq
\delta \equiv \mbox{det} (\phi_{\al\be}) \label{3.4}
\eeq
and
\beq
\ga \equiv - \frac{1}{D-2} \, .\label{3.5}
\eeq

It is straightforward to insert the decomposition (\ref{3.3})
 into the
Einstein-Hilbert action (\ref{3.1}) (see~\cite{5b} for details) and obtain
the following effective theory for the massless fields:
\bea
S & = & \int \di ^{D} x \sqrt{-g} \left\{
\frac{1}{2\kappa^2} R - \frac{1}{4} \delta^{-\ga}F^{\mu\nu\al}
F_{\mu\nu}^{\be}\phi_{\al\be} \right. \nnu \\
& & \left. +
\frac{1}{8\kappa^2}g^{\rho\lambda}\partial_{\lambda}\phi_{\al\be}
\partial_{\rho}\phi^{\al\be} - \frac{1}{8\kappa^{2}(D-2)}
g^{\rho\lambda}\partial_{\rho}\log \delta
\partial_{\lambda} \log \delta \right\} \, .\label{3.6}
\eea
Since we have compactified on tori, the vacuum expectation values of the
graviscalars are taken to be
\beq
\langle \phi_{\al\be} \rangle = \delta_{\al\be} \, .\label{3.7}
\eeq
The {\em diagonal}
 graviscalars---$\phi_{\al\al}$ for $\al =1,
 \ldots, E$---describe
fluctuations in the radii of the tori and the {\em off-diagonal}
ones describe fluctuations away from orthogonality of the compact
directions.

The determinant~(\ref{3.4}) describes variations of the volume of the
compact space. As such it is an effective space-time variation of the
$D$-dimensional Newton constant. However, the conformal rescaling
\beq
\h{g}^{\mu\nu} = \delta^{-\ga}g^{\mu\nu}\label{3.8}\, ,
\eeq
that was
introduced in (\ref{3.3}), ensures that the action for $g_{\mu\nu}$ is the
canonical one (as in (\ref{3.6})).  The volume factor $\delta^{-\ga}$
now appears as a
coupling to matter  instead of as a variation of
Newton's constant.

The graviphoton $A^{\al}_{\mu}$ is the gauge boson of the $U(1)$ group of
the rigid translations in the direction $\di y^{\al}$:
\beq
y^{\al} \rightarrow \tilde{y}^{\al} = y^{\al} -
\epsilon^{\al}(x)\label{3.9}\, ,
\eeq
and $F^{\al}_{\mu\nu}$ is the usual gauge-invariant field strength
\beq
F^{\al}_{\mu\nu} = \partial_{\mu}A^{\al}_{\nu} -
\partial_{\nu}A^{\al}_{\mu} \, .\label{3.10}
\eeq
Another way of looking at the graviphotons is
that they describe  fluctuations away from
orthogonality of the internal and external manifolds (see eq.~(\ref{3.3})).

\subsection{Coupling to massless matter and  propagators.}
\hspace*{2em}  In order to see anti-gravity  in $D$ dimensions
we have to
couple the theory
 to some  appropriately charged
 massive matter field. We do this by considering the
canonical coupling in $D+E$ dimensions of a massless scalar field
$\h{\Phi}$ to pure gravity
\beq
\h{S}_{\h{\Phi}} = -\frac{1}{2}
 \int \frac{\di ^{D+E} \hat{x}}{\rho (E)}
\sqrt{-\hat{g}}\: \h{g}^{\h{\mu}\h{\nu}} \partial_{\h{\mu}}
\h{\Phi}(\h{x}) \partial_{\h{\nu}}\h{\Phi}(\h{x}) + {\cal O} (\h{\Phi}^{3})  \,
{}.
\label{3.11}
\eeq
This is the generalization of the similar
analysis performed in ref.~\cite{Scherk} for the case $E=1$.

A mass $m_D$ is generated
in $D$ dimensions by giving the field  $\hat{\Phi}(\h{x})$ a momentum
$p_\al$ in the compact directions:
\beq
\h{\Phi}(\h{x}) = \Phi (x) e^{ip_{\al}y^{\al}} \, .\label{3.12}
\eeq
Up to a constant, $p_{\al}$ is the charge gauged by
the graviphoton
$A^{\al}_{\mu}$. The mass is simply
\beq
m_{D}^{2} = \delta^{\al\be}p_{\al}p_{\be} \, .
\eeq

The action (\ref{3.11}), once it has been
 decomposed into $D$ dimensions, describes the
coupling of $\h{\Phi}$ not only to the massless theory~(\ref{3.6}) but also
to the infinite tower of charged massive gravitons, graviphotons and
graviscalars that are obtained by giving a $y$-dependence to
$\h{g}_{\h{\mu}\h{\nu}}$.
Since we are only interested in long-distance physics we  can
safely restrict the
$y$-dependence to the field  $\h{\Phi}(\h{x})$. Integration over the internal
coordinates $y$'s then yields  the conservation of the charges:
\beq
p_{\al}^{in} + p_{\al}^{out} = 0 \, .
\eeq

It is straightforward to rewrite (\ref{3.11}) in terms of the
decomposition~(\ref{3.3}) and the result is the action
\bea
S_{\Phi} & = & -\frac{1}{2}
 \int \di ^{D} x
\sqrt{-g} \left\{
g^{\mu\nu} \left( \partial_{\mu} + i\sqrt{2}\kappa  p_{\al}^{in}
A^{\al}_{\mu} \right)
\Phi  \left( \partial_{\nu} + i\sqrt{2}\kappa  p_{\be}^{out}
A^{\be}_{\nu} \right) \Phi \right. \nnu \\
& & \left. -\phi^{\al\be}(p_{\al}^{in}
\Phi)(p_{\be}^{out} \Phi) \delta^{\ga}
 \right\} \, .\label{3.13}
\eea
If we insert the expansions
\bea
\sqrt{-g}g^{\mu\nu} & \equiv & \eta^{\mu\nu} + 2\kappa \tilde{\phi}^{\mu\nu}
\nnu \\
\phi_{\al\be}  & \equiv & \delta_{\al\be} + 2\kappa h_{\al\be}\label{3.14}
\eea
into (\ref{3.13}) we obtain a quadratic part describing the canonical
propagation of a massive scalar field $\Phi (x)$ and several cubic parts
corresponding  to the following
Feynman rules (in imaginary time):
\vspace*{1cm}
\beq
\hspace*{5cm} \kappa \left\{
p^{1}_{\mu}p^{2}_{\nu} + p^{1}_{\nu}p^{2}_{\mu}
-\frac{2}{D-2}\eta_{\mu\nu}m_{D}^{2} \right\} \label{A}
\eeq
\vspace*{1cm}
\beq
\hspace*{5cm} q_{\al} \left( p_{2}^{\mu} - p_{1}^{\mu} \right)  \label{B}
\eeq
\vspace*{1cm}
\beq
\hspace*{5cm} \frac{1}{\kappa} \left(
q_{\alpha}q_{\beta} + \frac{2}{D-2} \kappa^2
m^{2}_{D} \delta_{\alpha\beta} \right) \, .   \label{C}
\eeq
\vspace*{1cm}
Here the solid lines represent the massive state with charges
$q_{\al}$
($\al = 1, \ldots, E$) given by
\beq
q_{\al} = \sqrt{2}\kappa p_{\al} \, . \label{3.15}
\eeq
We see that the graviscalar couples to the matrix of the charges
$q_{\al}q_{\be}$, and, because of the rescaling~(\ref{3.8}), also to
the trace of the energy-momentum tensor.

The propagators of graviton, graviphoton and graviscalars can be extracted
from~(\ref{3.6}) and are given by
\bea
\lefteqn{ \langle \g^{\mu_1 \nu_1} (p) \g^{\mu_2 \nu_2} (-p)
\rangle =  } \nnu \\
& & \frac{1}{2p^2} \left\{ \eta^{\mu_1 \mu_2} \eta^{\nu_1 \nu_2} +
\eta^{\mu_1 \nu_2}\eta^{\nu_1 \mu_2} - \eta^{\mu_1 \nu_1}
\eta^{\mu_2 \nu_2} \right\} \label{A.1}\, ,
\eea
\beq
\langle  A_{\mu}^{\al} (p) A_{\nu}^{\be} (-p) \rangle
=  \frac{1}{p^2}
\delta^{\al \be} \eta_{\mu \nu}  \label{A.2}
\eeq
and
\bea
\lefteqn{  \langle
h_{\al_1 \be_1} (p) h_{\al_2 \be_2} (-p)
\rangle = } \nnu \\
& & \frac{1}{2p^2} \left\{ \delta_{\al_1 \al_2} \delta_{\be_1 \be_2}
+ \delta_{\al_1 \be_2} \delta_{\be_1 \al_2}
- \frac{2}{D+E-2} \delta_{\al_1 \be_1} \delta_{\al_2 \be_2} \right\}
\label{A.3}\, ,
\eea
respectively.
We work in De Donder gauge
\beq
\partial_{\nu} \g^{\mu \nu} = 0 \label{A.4}
\eeq
for the graviton and Lorentz gauge
\beq
\partial_{\mu} A^{\mu}_{\al} = 0 \label{A.5}
\eeq
for the graviphotons.

\subsection{Anti-gravity at work.}
\hspace*{2em}  It is now possible to compute
in the Born approximation the
elastic four-point amplitude for the
static exchange between
two charged massive states.  It  comes from the sum of
the three Feynman diagrams in fig.1. For simplicity we assume that
the two external states have only one  non-zero charge
\beq
q_{i} = \sqrt{2} \kappa \epsilon_{i}m_{i}
 \:\:\:\: \epsilon_{i}=\pm 1
\:\: \: i=1,2 \, ,
\eeq
corresponding to a compact momentum that points entirely in one
direction,  for example $\di y^1$.

The static limit yields the following amplitude:
\bea
A(1234) & = & - \frac{4\kappa^2}{t} m_{1}^{2}m_{2}^{2}
\left\{ \frac{D-3}{D-2} + \frac{1}{D-2} + (\epsilon_{1}\epsilon_{2})^2
- 2\epsilon_{1}\epsilon_{2} \right\} \nnu \\
& = &  - \frac{4\kappa^2}{t} m_{1}^{2}m_{2}^{2} \left( 1
-\epsilon_{1}\epsilon_{2} \right)^2 \label{3.16} \, .
\eea
We have displayed in the first line of eq.~(\ref{3.16}) the different
contributions. The first one comes from
 the graviton exchange. Notice that
this  term cancels in $D=3$, where there is no physical graviton. The second
and third one are due to the graviscalar $\phi_{11}$, and
the last one comes from the exchange
of the graviphoton $A^{1}_{\mu}$.

The amplitude turns out to be proportional to
$(1 - \epsilon_{1}\epsilon_{2} )^2$ and therefore
we witness  anti-gravity at work:
the static potential vanishes between like charges and is enhanced
between opposite charges.

The above balance will be upset if (in the Born
approximation) the external
states can interact in $D+E$ dimensions by
other means beside pure gravity. We would then have to add extra Feynman
diagrams to the ones depicted in fig.1.
\begin{figure}
\vspace{6cm}
\caption{The three Feynman diagrams contributing to the static
potential of two massive charged states.}
\end{figure}
To prevent this from occurring, we can always
 use an excited mode
of the $(D+E)$-dimensional graviton itself  (see eq.~(\ref{2.8}))
as  external states.
This state is neutral with respect to all other gauge groups.

The cancellation (\ref{3.16}), which may seem rather mysterious at first,
has a very simple interpretation in $D+E$ dimensions~\cite{Scherk}.  The
amplitude~(\ref{3.16}) is  by construction
nothing but the
amplitude for  the gravitational scattering
of two massless particles in the special case where the
momenta are entirely in the compact direction $\di y^1$. If
$\epsilon_{1} = \epsilon_{2}$, the two particles move in the same
direction at the speed of light and will never meet, giving  a
vanishing amplitude (in the center-of-mass frame both carry zero energy).
If  $\epsilon_{1} = -\epsilon_{2}$, the two particles are colliding and
it is possible to have a non-zero amplitude.

In fact, if we consider the $(D+E)$-dimensional  four-point
 amplitude for the gravitational
 scattering of two massless states, which is given for $t
\rightarrow 0$ by
\beq
\h{A}(1234) = -\frac{2\h{\kappa}^2}{t} \left\{
\h{p}_{1}\cdot \h{p}_{4} \: \h{p}_{2}\cdot \h{p}_{3} +
\h{p}_{1}\cdot \h{p}_{3} \: \h{p}_{2}\cdot \h{p}_{4} -
\h{p}_{1}\cdot \h{p}_{2}  \:\h{p}_{3}\cdot \h{p}_{4} \right\}\, ,
\eeq
and evaluate it in the special kinematical situation
\bea
\h{p}_{1} & = & -\h{p}_{2}  = (m_1 , 0; p^{\al}_{1}) \nnu \\
\h{p}_{4} & = & -\h{p}_{3}  = (m_2 , 0; p^{\al}_{2})
\eea
which corresponds to
the static limit in $D$ dimensions, we obtain
\beq
A(1234) = - \frac{4\kappa^2}{t} m_{1}^{2}m_{2}^{2} \left( 1 -
\frac{\sum_{\al=1}^{E} p_{1}^{\al}p_{2}^{\al}}{m_1 m_2} \right) ^2\,
.\label{3.24}
\eeq
The amplitude~(\ref{3.24})
 is the generalization of eq.~(\ref{3.16}) to the case where all $E$
charges are non-zero.

\subsection{Anti-gravity and supersymmetry.}
\hspace*{2em} Although the anti-gravity mechanism we have presented
does not require supersymmetry, most known candidates for the
quantum theory of gravity are endowed with
some kind of space-time supersymmetry.

If the $(D+E)$-dimensional theory is supersymmetric, the
$D$-dimensional action~(\ref{3.6}) will be part of an extended supergravity
theory. In these cases  it is possible to give an elegant
characterization of the charged massive states that give rise to
anti-gravity, namely,
they fall into massive multiplets with all central
charges in the supersymmetry algebra equal and given by
$Z= \pm 2m$. This may be seen as follows.  In $D+E$
dimensions the states
we are considering
are massless and fall into a massless multiplet of
the $D+E$ dimensional supersymmetric algebra with no central charge
(being massless there is no way to provide a dimensionful central
charge). On compactification the $D+E$ dimensional supersymmetric algebra
is reduced to the extended $D$ dimensional supersymmetric algebra with
the compact momenta playing the role of central charges. The charged
massive copies of the graviton multiplet will have all central charges
$Z= \pm 2m$ simply because this is the only way to obtain a massive
representation with the same dimension of the massless graviton multiplet
\cite{3b}.

\subsection{Section 2 revisited.}
\hspace*{2em} Let us now return to the three examples of anti-gravity
presented in section 2 to point out how they fit into the general
framework of the present section.

For the type-II superstring (compactified on tori) the situation is
clear: if the state~(\ref{2.8}) has
at least one compactified momentum, it will be  exactly of the
charged type  we have
considered here. Although in ten dimensions
we have
not only the gravitational field but also the dilaton and the rank two
anti-symmetric tensor, neither of these give rise to diagrams of the type shown
in fig.1 for the on-shell massless states we have
considered\footnote{The same is true for the
vector field and the rank three anti-symmetric tensor field that describe
the massless states in the $R-\overline{R}$ sector of the superstring.}.
Hence, anti-gravity is an exact feature of the four dimensional
low-energy effective field theory, which is $N=8$, $D=4$  supergravity.

The formulation in terms of  central charges is also clearly
exhibited by the string example. The charged states of the type
(\ref{2.8}) are
essentially  massive copies of the states in the massless multiplet.
In general, those having a momentum
$p_{\al}$ in the compact direction $\al \, (\al = 1, \ldots, 6)$ fall into
a massless multiplet of the N=8 supersymmetry algebra with all four central
charges equal to
\beq
Z = 2 p_{\al} = \pm 2m \label{2.5}
\eeq
as it can be understood by rewriting the
ten-dimensional supersymmetry algebra
of the superstring
(which does not contain a central charge) in the four-dimensional language.
Precisely because of (\ref{2.5}) these multiplets have the same dimension
as the massless one (which is 256).

On the contrary, the massive string excitations with mass $m_{10}$
fall into massive representations
of the $N_+ = N_- = 1$ supersymmetry algebra. On compactification they too
will have a central charge $Z = 2p_{\al}$ but since the 4-dimensional
mass is now
\beq
m_4 = \sqrt{(p_{\al})^2 + m_{10}^2}   \label{2.5a}\, ,
\eeq
the central charge is not equal to $\pm 2m_{4}$ and the multiplets
will retain their higher dimensionality (which for the case of the scalar
multiplet describing the first excited level of the string  is
$256 \times 256$).

In the model proposed by Scherk, the generalized dimensional reduction
breaks the $N=8$ supersymmetry completely and hence gives
a mass to most of
the 256 states in the $D=4$ graviton multiplet. By construction, all these
states are charged under the $U(1)$ group
of the fifth dimension (see
eq.(\ref{2.3})). Since the
corresponding gravivector $A^{1}_{\mu}$ and graviscalar $\phi_{11}$ remain
massless (as does, of course, the graviton) the balance of forces
in~(\ref{3.16}) remains intact.

This example shows that anti-gravity need not be a phenomenon restricted
to very massive states coming from an
excited momentum in one of the compact  directions.
Ordinary matter could, therefore, carry graviphoton charges.
Unfortunately, the
question whether it actually does will not be answered until
we have a better understanding of how masses are generated in nature.

The $N=2$ model of Zachos is somewhat a special case. It contains
only the graviton multiplet (that is, the graviton, two gravitinos and a
single graviphoton) and a massive multiplet with central charge $Z=2m$.
There are no graviscalars.  Even though a theory of $N=2$
supergravity in $D=4$ may be obtained by dimensional reduction of $N_+ =
N_ - =1$ supergravity in six dimensions, this will contain further massless
multiplets in which the second graviphoton and three graviscalars can be
found. We see that the $N=2$ model of Zachos cannot be obtained by
compactification---the problem being that the $N=2$ multiplets are too
small for all graviphotons and graviscalars to fit inside.

\section{How strong is anti-gravity?}
\setcounter{equation}{0}
\hspace*{2em} As we have defined it, anti-gravity
is just the cancellation of the Newtonian part of the static potential.
It is natural to ask what happens if we take non-linear corrections into
account. In this section we investigate this question.

\subsection{Non-linear corrections.}
\hspace*{2em} The $N=2$, $D=4$ model of Zachos is easily dealt with by
noting that the bosonic part of the massless theory is just gravity
coupled to a single U(1) gauge field. The exact solution generated by a
point-mass with charge $Q$
is the Reissner-Nordstr\o m space-time geometry~\cite{RN}
\bea
\di s^2 &=& - \left( 1 - \frac{2MG_N}{r} + \frac{G_N Q^2}{4\pi r^2} \right)
\di t^2
+ \left( 1 - \frac{2MG_N}{r} + \frac{G_N Q^2}{4\pi r^2} \right) ^{-1} \
\di r^2 \nnu \\
& & + r^2 \left( \di \theta^2 +\sin ^{2}\theta
 \di \phi^2 \right) \label{4.1}
\eea
 The solution of the corresponding Maxwell equations
gives a static Coulomb potential:
\beq
A^0 = -\frac{Q}{4\pi r} \:\:\:\: ; \: A^i = 0 \, .
\eeq
The value of the charge  for which anti-gravity occurs is
\beq
Q=\frac{\sqrt{2}}{2}\kappa M .
\eeq
However, it is clear that the ensuing cancellation of forces in the
static limit is only a feature of the linear approximation; thereby the
theory is
an example of weak anti-gravity. Notice that~(\ref{4.1}) contains  a horizon
(For a discussion of the Reissner-Nordstr\o m space-time in
this particular case, see~\cite{Carter}).

The $N=2$ example strongly
suggests that weak anti-gravity is associated to the case
in which the theory cannot be derived by compactification.
On the other hand,
the $N=8$, $D=4$ model of the other two examples
of section 2 belongs to the more
general situation of section 3---where anti-gravity arise
as a re\-sult of com\-pac\-ti\-fi\-ca\-tion---and, as
 we shall see,
 represents  strong anti-gravity. In this case
we expect no horizon
to appear. The reason for this is simple. The charged massive state
generating the gravitational field in four dimensions is nothing but a
massless state moving in  one of the compact directions. The exact
solution of the metric generated by such a particle is well-known and
contains no horizons. It is the Aichelburg-Sexl (AS)
 metric~\cite{AS}.

\subsection{The exact solution based on the generalized AS metric.}
\hspace*{2em}  The Aichelburg-Sexl metric is the solution of
Einstein's equations for a point source moving at the speed of light. It
can be
obtained by an infinite boost of the Schwarzschild solution and
in $D=4$ is given by:
\beq
\di s^2 = -\di u \di v - (8G_N E\log r )\delta(u) \di u^2 +
 \di x_{\perp}^{2} \label{AS}\, ,
\eeq
where $u=t-y$, $v=t+y$ and $x_{\perp}$ is the vector of the components that
are transverse with respect to $\di y$, the direction of motion of the
massless particle. $E$ is the energy of the particle.

What we will actually need is a straightforward generalization of
(\ref{AS}) to include an arbitrary number of space-time dimensions and a
more general energy profile.  The solution becomes
\beq
\di s^2 = -\di t^2 + \di y^2  + \di \h{x}_{\perp}^{2}
+ f(\h{x}_{\perp}, t-y ) (\di t  - \di y )^{2} \, .
  \label{4.5}
\eeq
Here $y$ can be any one of the $E$ compact coordinates.
 The metric (\ref{4.5}) is the most general solution that one can write. It
 describes the propagation in the  direction $\di y$ of a
 shock wave with a shape specified by the function
 $f(\h{x}_{\perp}, t-y)$,
 which is related
 to the energy profile $\h{\rho} (\h{x}_{\perp}, t-y ) $
  by  the Einstein equations
\beq
\nabla^{2}_{\h{x}_{\perp}} f(\h{x}_{\perp}, t-y) =
 -16\pi \h{G}_N \h{\rho} (\h{x}_{\perp}, t-y) \, ,\label{4.6}
\eeq
where  $\nabla^{2}_{\h{x}_{\perp}} $ is the flat-space Laplacian in the
transverse coordinates. If we take
$f(\h{x}_{\perp}, t-y)  = f(\h{x}_{\perp})\delta(t-y)$,
eq.~(\ref{4.6}) describes the shock wave due to a
source completely localized in the beam direction~\cite{venezia}.
The shock wave reaches
out to infinity in the transverse directions.

The fact that Einstein equations take the linear
form~(\ref{4.6}) is a remarkable feature of the shock wave solution.
It means that we can superpose individual solutions to create any kind of
profile in the beam direction. In particular, we can choose
 $f(\h{x}_{\perp}, t-y) $ to be independent of
  $t-y$. This
 corresponds to a wave that is completely smeared out in the compact
 direction $\di y$. We likewise smear out the wave in the transverse
 compact  directions by taking  $f(\h{x}_{\perp}, t-y) $
 to be a function of
 $\vec{x}$ only, the $D-1$ non-compact spatial coordinates.

 If we now  take the energy profile to be
 \beq
\h{\rho} (\vec{x}) = M \frac{1}{\rho(E)}\delta^{D-1} (\vec{x})
 \eeq
 this  will
 correspond in $D$ dimensions to having a point-like source at the
 origin with mass $M$ and charge $q = \sqrt{2}\kappa M$. Accordingly,
 $f(\vec{x})$ is the spherically symmetric solution of Laplace
  equation in $D-1$ dimensions:
  \beq
  \sum_{i=1}^{D-1} \partial_i \partial_i f(\vec{x}) = -16\pi G_N M
  \delta^{D-1}(\vec{x} ) \, ,
  \eeq
  that is, in terms of the radial coordinate
  $r = \left( \sum_{i=1}^{D-1}x_{i}^{2}
  \right) ^{1/2}$,
  \beq
  f(r) = \frac{2\kappa^2 Mr^{3-D}}{\Omega_{D-1}(D-3)}\label{4.9} \, ,
  \eeq
  where
  \beq
  \Omega_D = 2 \frac{\pi^{D/2}}{\Gamma (D/2)}
  \eeq
  is the solid angle in $D$ dimensions.

  What we have obtained is an exact solution of the Einstein theory in
  $D+E$ dimensions  for a  fluid of massless particles
  moving in the direction $\di y$
  with a total momentum $p$. The
  fluid is homogeneously smeared out in all compact coordinates (hence
  our solution satisfies all periodicity requirements) but is localized
  at the origin in the $D-1$ dimensional non-compact space.
  A $D$-dimensional observer is living  so-to-speak
  inside the shock wave at all
  times.

  By construction, then, we have also found the exact solution of the the
  non-linear theory~(\ref{3.6}) corresponding to a charged point mass at
  rest at the origin. In fact,
 the solution~(\ref{4.5}), (\ref{4.9}) can be recasted in $D$-dimensional
   language by using the
  decomposition~(\ref{3.3}). Taking the compact momentum in the
direction $\di y^1$ one finds
  \bea
  \di s^2 & = & \left( 1+f(r) \right) ^{-\ga}
  \left[ -\left( 1+f(r) \right) ^{-1} \di t^2 + {\di \vec{x}}^{2} \right]
  \nnu \\
  A^{\al}_{\mu} & = & \frac{\epsilon}{\sqrt{2}\kappa}
   \frac{f(r)}{1+f(r)}
  \delta^{0}_{\mu}\delta^{\al}_{1} \nnu \\
  \phi_{\al\be} & = &  \delta_{\al\be} + f(r)\delta^{1}_{\al}\delta^{1}_{\be}
  \label{4.11}\, .
  \eea
Here $\epsilon$ is the sign of the charge $q_1$.

  The simplicity of the solution~(\ref{4.5}),(\ref{4.9}) should be contrasted
to
  the more complicated~(\ref{4.11}).
  The $D$-dimensional field equations of which~(\ref{4.11}) is a
  solution are quite complicated  too. We give them in  the appendix.
Because only the fields $\phi_{11}$ and $A_{\mu}^1$ are
excited in (\ref{4.11}), the action
(\ref{3.6})  reduces in this case to the simpler one
\bea
S & = & \int \di ^D x \sqrt{-g} \left\{ \frac{1}{2\kappa^2} R
- \frac{1}{4} \exp ( \frac{D-1}{D-2} \log \phi ) (F_{\mu \nu})^2
\right. \nnu \\
 & & \left. - \frac{1}{8\kappa^2} \frac{D-1}{D-2} (\partial_{\mu} \log \phi )^2
\right\}  \label{minaction} \, ,
\eea
where $F_{\mu \nu} \equiv F_{\mu \nu}^1$ and $ \phi \equiv \phi_{11}$.
The fields~(\ref{4.11})
are  a solution of the field equations derived from this
action.

 Various theories of gravity coupled to scalar and Maxwell fields have
 been considered in the
literature~\cite{Gibbons,Chodos,Horowitz} of
which the action (\ref{minaction}) and our solution (\ref{4.11})
 are a special case. The idea that the solution could be obtained
 by boosting the Schwarzschild solution into a compact fifth direction
 was pointed out in~\cite{Gibbons}.

  It should be noted that if we take $D+E=10$ and $D=4$ the
theory~(\ref{3.6}) is part of the four-dimensional Lagrangian for
$N=8$ (or alternatively $N=4$) supergravity.
Therefore, what we have found is also a
   solution to the field equations of $N=8$ (or $N=4$)
  supergravity in the case where only
  one of the graviphoton charges is
  nonzero and given by eq.~(\ref{3.15}).
 The other fields that are present in
the $N=2$ ($N=1$) supergravity
Lagrangian in ten dimensions, from which the $N=8$ ($N=4$) theory is
obtained by dimensional reduction~\cite{13b,13c},
can be consistently put to zero in the ten dimensional equations
of motion---as one can check by inspection. A similar observation was
made in~\cite{Gibbons}.

\subsection{Discussion of the exact solution.}
  \hspace*{2em} To study the non-linear corrections to anti-gravity we
  now consider the
  motion of a test particle of
mass $m_D$ and charge $\epsilon ' \sqrt{2} \kappa m_D$
   in the above solution. From the $D$-dimensional point of
  view the equations of motion can be obtained by requiring covariant
  energy-momentum conservation of the combined system (test particle + field).
  This can be done, and the resulting equations are given in the
  appendix.

It is, however, much easier to take the $D+E$-dimensional point of view:
the test particle is massless with compact momentum $p_1$ and its
equations of motion are just the geodesic equations of motion in the
shock-wave metric given by (\ref{4.5}) and (\ref{4.9}). These are
easily obtained and reads:
\bea
\dot{t} & = & \frac{E}{m_D} \left(1+ f(r) \right)
- \epsilon \epsilon' f(r) \nnu \\
\dot{y} & = & \frac{E}{m_D} \epsilon f(r) + \epsilon' (1-f(r)) \nnu \\
\dot{\theta} & = & \frac{L}{(m_D r^2)} \nnu
\eea
\beq
\half \dot{r}^2 - \half \frac{E^2}{m_D^2} \left(1- \epsilon \epsilon'
\frac{m_D}{E} \right) ^2
f(r) + \frac{L^2}{2 m_D^2 r^2} = \half \left( \frac{E^2}{m_D^2}
-1 \right) \label{4.12} \, .
\eeq
In eq.~(\ref{4.12})
the transverse compact directions decouple completely. We have also
denoted
$y \equiv y^1$.
$E$ is the energy of the test particle in $D$ dimensions and $L$  is
its angular
momentum.
 The motion
is planar in $D$ dimensions and can be
 described by introducing polar coordinates
$(r,\theta)$ in that plane. The dot denotes differentiation with respect to
the proper time $\tau$. In the case of a massless test particle,
where $m_D =
\epsilon' = 0$, it is necessary to introduce the rescaled parameter
$\sigma = \tau/m_D$. Alternatively one may take the limit $E/m_D \gg 1$
in the final results.

The canonical momentum in the $\di y$-direction is
\beq
\pi = \h{g}_{y t} \dot{t} + \h{g}_{y y} \dot{y} = m_D \epsilon'
\label{4.14}\, .
\eeq

By differentiating the radial equations of motion in (\ref{4.12})
 with respect to $\tau$
we obtain (in the case $L=0$ of radial motion):
\beq
\ddot{r} = \half \frac{E^2}{m_D^2} \left(
1-\epsilon \epsilon' \frac{m_D}{E} \right) ^2
f'(r)  \label{4.15}\, .
\eeq
The right-hand side is either negative or zero. In the static limit $E=m_D$
we recover anti-gravity:
\beq
\ddot{r} = \half (1-\epsilon \epsilon')^2 f'(r)  \label{4.16}
\eeq
---that is, the vanishing of the radial acceleration for like charges.
 It is  now
an exact result that holds to all orders! Even arbitrarily close
to the singularity at $r=0$ the test particle remains at rest when
$\epsilon = \epsilon'$. There is no horizon, as  it is also clear from the
form (\ref{4.11}) of the metric: $g_{tt}$ is negative everywhere.

However, as soon as the particle starts  moving in the radial direction,
$E/m_D > 1$ and gravity becomes slightly stronger than the other forces
with the result that the acceleration becomes non-zero and towards the
point-mass.

The entire problem of motion contained in (\ref{4.12}) is in fact
mathematically equivalent to the Kepler problem with an effective potential
\beq
\phi^N_{eff} (r) = - \half \frac{E^2}{m_D^2}
\left( 1-\epsilon \epsilon' \frac{m_D}{E}
\right) ^2 f(r)
 \label{4.17}
\eeq
and an effective Newtonian energy (kinetic + potential) which is
\beq
E^N_{eff} = \half m_D \left[ \left( \frac{E}{m_D}
\right) ^ 2 - 1 \right]   \label{4.18}\, .
\eeq
Therefore, in $D=4$ the test particle trajectory will be elliptic,
parabolic or hyperbolic in the $(r,\theta)$-coordinate plane, depending
on whether $E/m_D$ is less than, equal to or greater than one. In the
latter case, we can compute the deflection angle  $\Delta \varphi$
to all orders in $G_N$. The result is
\beq
\sin \frac{\Delta \varphi}{2} = \frac{z}{\sqrt{1+z^2}}
  \hspace{2cm} (D=4)  \label{4.19}\, ,
\eeq
where
\beq
z =
\frac{2MG_N}{b} {
\frac{(1-\epsilon \epsilon' \frac{
\displaystyle m_D}{\displaystyle E})^2}{1- \left(
\frac{\displaystyle m_D}
{\displaystyle E} \right)^2 }}  \hspace{2cm} (D=4)   \label{4.20}
\eeq
and the impact parameter $b$ is defined by $L^2 = b^2 (E^2 - m^{2}_D )$.
For a neutral massless test particle (or for a very energetic charged one)
we find
\beq
z = \frac{2MG_N}{b}  \hspace{2cm}  (D=4)   \label{4.21} \, ,
\eeq
and in the limit $b \rightarrow \infty$ we recover  Einstein's formula
for the deflection of light:
\beq
\Delta \varphi = \frac{4G_N M}{b}   \hspace{2cm}  (D=4)  \, .  \label{4.22}
\eeq
Notice the absence in (\ref{4.19}) of a term proportional to $G_{N}^2$,
a characteristic feature of scattering in a shock-wave metric~\cite{venezia}.

The  discussion above shows
that life inside the shock wave is so simple that
the fully relativistic theory  is effectively Newtonian.
The reason for this remarkable fact lies in the linearity of the full
Einstein equations on the class of shock wave metrics.

\section{Anti-gravity in general.}
\setcounter{equation}{0}
\hspace*{2em}  In the previous section we have established the presence
of strong
anti-gravity in any theory of gravity that is obtained from a
higher-dimensional theory by toroidal compactification.
In this section we consider
what happens if we try to relax the condition
that the compact dimensions should be tori.

\subsection{Propagation in compactified Einstein theory.}
\hspace*{2em}  We
consider again the Einstein-Hilbert action (\ref{3.1}) in $D+E$
dimensions. But now we take a  background more general than (\ref{3.7}),
namely a manifold of the form $M_D \times K$,
where $M_D$ is flat $D$-dimensional
Minkowski space-time and $K$ is some general compact $E$-dimensional
manifold of volume $\rho(E)$.

We may still decompose the $D+E$-dimensional metric into gravitational,
gravivectorial and graviscalar fields along the lines of
 eq.(\ref{3.3}).
Again different $D$-dimensional particles can be identified with different
modes on the compact manifold through the ansatz:
\bea
\g^{\mu \nu} (x,y)  & = & \g^{\mu \nu} (x)  u (y) \nnu \\
A_{\mu}^{\beta} (x,y) & = & A_{\mu} (x) u^{\beta} (y) \nnu \\
h^{\alpha \beta} (x,y) & = & h (x) u^{\alpha \beta} (y) \label{5.1}\, .
\eea
On the flat manifold $(S^{1})^E$ made of tori,
 the mass as seen in $D$ dimensions was
simply the eigenvalue of the Laplacian in $E$ dimensions. Corresponding to
the one scalar, $E$ vectorial and $E(E+1)/2$ tensorial zero modes we had
a similar number of massless gravitons, graviphotons and graviscalars.

On a curved manifold  we still have a unique scalar zero mode---the one
that is
constant---that
gives in $D$ dimensions  a massless graviton. On the other hand,
 for the graviphoton
and the graviscalar the
situation is more complicated.  Their
mass is now given by the eigenvalue of a more general operator
that involves not only the Laplacian but also the coupling of these modes
to the background curvature. Furthermore, the concept of a constant
mode becomes non-trivial for  vectors and tensors because we must
replace the ordinary derivatives  by the covariant ones.

Our aim is to identify those compact manifolds $K$ for which the
anti-gravity effect persist, that is, for which one graviphoton and one
graviscalar remain massless.

To analyze this question we con\-si\-der the pro\-pa\-ga\-ting
part of the
Ein\-stein-Hil\-bert action (\ref{3.1}). We ex\-pand the
$D+E$-di\-men\-sio\-nal met\-ric
(\ref{3.3})
in the fluc\-tua\-tions $(\g_{\mu \nu}, A_{\mu}^{\al},h_{\al \be})$
where, in analogy with (\ref{3.14}), we take
\bea
\sqrt{-g} g^{\mu \nu} & \equiv & \eta^{\mu \nu} + 2\kappa \g^{\mu \nu}
\nnu \\
\phi_{\al \be} & \equiv & g^{(B)}_{\al \be} + 2\kappa h_{\al \be}
\label{5.2}\, .
\eea
By inserting these expansions into eq.(\ref{3.3}) we find
\beq
\h{g}_{\h{\mu} \h{\nu}} = \h{g}^{(B)}_{\h{\mu} \h{\nu}} + \h{h}_{\h{\mu}
\h{\nu}}  \label{5.5}\, ,
\eeq
where the background metric (after a conformal rescaling of the
coordinates in Minkowski space) is given by
\beq
\h{g}^{(B)}_{\h{\mu} \h{\nu}} = \left( \begin{array}{cc}
\eta_{\mu \nu} & 0 \\ 0 & g^{(B)}_{\alpha \beta} \end{array} \right)
\label{5.6}
\eeq
and the fluctuation is parametrized as follows:
\bea
\h{h}_{\mu \nu} & = &
2\kappa \left\{ -\g_{\mu \nu} + \frac{1}{D-2}
(\g - h) \eta_{\mu \nu} \right\}
+ 2 \kappa^2 A_{\mu}^{\alpha} A_{\nu \alpha} \nnu \\
& & + 2\kappa^2 \left\{ -\frac{1}{D-2} (\g_{\rho \sigma})^2 \eta_{\mu \nu}
+ \frac{1}{(D-2)^2} \g^2 \eta_{\mu \nu} + 2\g_{\mu \rho} \g^{\rho}_{\nu}
- \frac{2}{D-2} \g \g_{\mu \nu} \right\} \nnu \\
& & + 2\kappa^2 \left\{ \frac{1}{D-2}
 (h_{\alpha \beta})^2
\eta_{\mu \nu} + \frac{1}{(D-2)^2} h^2
\eta_{\mu \nu} \right\}  \nnu  \\
& & -
\frac{4\kappa^2}{D-2} h (-\g_{\mu \nu} +
\frac{1}{D-2} \g \eta_{\mu \nu})  +  (\mbox{higher order}) \nnu \\
\h{h}_{\mu \beta} & = & -\sqrt{2} \kappa A_{\mu \beta} - 2\sqrt{2} \kappa^2
A_{\mu}^{\alpha} h_{\alpha \beta} \nnu \\
\h{h}_{\alpha \beta} & = & 2\kappa h_{\alpha \beta} \label{5.7}\, .
\eea
In this section all external indices are raised
and lowered with the Minkowski metric and all internal indices with the
internal background metric $g^{(B)}_{\al\be}$. $\g$ and $h$ denote the trace
of $\g_{\mu \nu}$ and $h_{\al \be}$, respectively.
It is now  straightforward to expand the action (\ref{3.1}) in
powers of the fluctuation $\h{h}$. One obtains~\cite{MTW}
\beq
\h{S} = \h{S}^{(0)} + \h{S}^{(1)} + \h{S}^{(2)} + \ldots \label{5.8}
\eeq
with
\bea
\h{S}^{(0)} & = & \frac{1}{2\kappa^2} \int \frac{\di^{D+E} \h{x}}{\rho(E)}
\sqrt{-\h{g}^{(B)}} \h{R}^{(B)} \nnu \\
\h{S}^{(1)} & = & \frac{1}{2\kappa^2} \int \frac{\di^{D+E} \h{x}}{\rho(E)}
\sqrt{-\h{g}^{(B)}} \h{h}^{\h{\mu} \h{\nu}}
\left( - \h{R}^{(B)}_{\h{\mu} \h{\nu}} + \frac{1}{2} \h{g}^{(B)}_{\h{\mu}
\h{\nu}} \h{R}^{(B)} \right)
\nnu \\
\h{S}^{(2)} & = & \frac{1}{2\kappa^2}  \int \frac{ \di^{D+E} \h{x}}{\rho(E)}
 \sqrt{-\h{g}^{(B)}} \left\{
\left( \frac{1}{8} \h{h}^2 - \frac{1}{4} \h{h}^{\h{\mu} \h{\nu}}
\h{h}_{\h{\mu} \h{\nu}} \right)
\h{R}^{(B)}
+ \h{h}^{\h{\mu}}_{\h{\rho}} \h{h}^{\h{\rho} \h{\nu}} \h{R}^{(B)}_{\h{\mu}
\h{\nu}} \right.
\nnu \\
& &  -\frac{1}{2} \h{h} \h{h}^{\h{\mu} \h{\nu}}
\h{R}^{(B)}_{\h{\mu} \h{\nu}} + \frac{1}{4} \nabla_{\h{\mu}} \h{h}
\nabla^{\h{\mu}} \h{h} -\frac{1}{2}
\nabla_{\h{\mu}} \h{h} \nabla_{\h{\nu}} \h{h}^{\h{\mu} \h{\nu}}  \label{5.9} \\
& & \left.
 + \frac{1}{2} \nabla_{\h{\rho}}
\h{h}^{\h{\nu}}_{\h{\mu}} \nabla_{\h{\nu}} \h{h}^{\h{\rho} \h{\mu}} -
\frac{1}{4} \nabla_{\h{\rho}}
\h{h}^{\h{\nu}}_{\h{\mu}} \nabla^{\h{\rho}} \h{h}^{\h{\mu}}_{\h{\nu}}
\right\} \nnu\, .
\eea
The covariant derivatives are with respect to the background metric.
Similarly $\h{h}$ is the trace with respect to $\h{g}^{(B)}$.
Since we have considered only pure gravity in $D+E$ dimensions, tadpole
terms
linear in the fluctuation will appear unless the background metric
satisfies Einstein equations in the vacuum, i.e. unless the manifold
$K$ is Ricci flat:
\beq
R^{(B)}_{\alpha \beta} = 0  \label{5.10}\, .
\eeq
Even though we could
imagine
creating any kind of background metric by introducing
appropriate classical sources, for the moment, we proceed and
consider the
part of the Einstein-Hilbert action quadratic in the fluctuations
$(\g,A,h)$ without
making any
assumptions on the background curvature. We come back to this point at
the end of the section.

\subsection{The graviton.}
\hspace*{2em}
Let us first consider the rather simple case of the graviton.  All terms
involving derivatives with respect to the coordinates on the compact
space vanish and we are left with the following few terms:
\bea
S_{\g}^{(2)} & = &
\frac{1}{2} \int \di^Dx
\left\{ 2 \nabla_{\mu}
\g^{\nu}_{\rho} \nabla_{\nu} \g^{\mu \rho} - (\nabla_{\mu} \g^{\nu \rho})^2
+ \frac{1}{D-2} (\nabla_{\mu} \g)^2  \right\} \nnu \\
& & - \Lambda \int  \di^Dx
\left\{ \frac{2}{(D-2)^2} \g^2 - \frac{2}{D-2} (\g_{\mu \nu})^2
\right\} \label{5.11}\, .
\eea
The compact space  variables
have been integrated out  (with the zero mode $u$ of eq.~(\ref{5.1})
being normalized to unity). The first three terms give simply the
propagator of the graviton in $D$ dimensions after an appropriate
gauge-fixing term is added to fix the invariance under external
reparametrizations $x^{\mu} \rightarrow x^{\mu} - \epsilon^{\mu}(x)$.

The last terms in eq.(\ref{5.11}) represent
the coupling of the graviton to a cosmological constant
\beq
\Lambda = -\half \int \frac{\di^Ey}{\rho(E)} \sqrt{g^{(B)}} R^{(B)}
\label{5.12}
\eeq
that is produced by the curvature of the internal space.  If we
assume that the background satisfies the Einstein  equations in vacuum,
this
cosmological constant vanishes.

\subsection{Conditions for a massless graviphoton.}
\hspace*{2em}
 Next we turn our attention to the graviphoton. The part of the
 action quadratic in the graviphoton field can be written as
\beq
S_{A}^{(2)} =  \half \int \di^Dx
 \left\{  \n_{\mu} A^{\nu} \n_{\nu} A^{\mu } -
 (\n_{\mu} A^{\nu })^2  -  m_V^2  A^{\mu} A_{\mu} \right\} \, ,
\label{5.15}
\eeq
where the mass $m_V^2 $ is given by
\bea
m_V^2  & =  & - \int \frac{\di^Ey}{\rho(E)} \sqrt{g^{(B)}} \left\{
 2u^{\al} u^{\be} R^{(B)}_{\al \be} +
 \n_{\ga} u^{\be}
 \n_{\be} u^{\ga} - (\n_{\ga} u^{\al})^2
 \right\}  \label{5.16}\, .
 \eea
To arrive at (\ref{5.15}) and (\ref{5.16})
we have introduced  the ansatz (\ref{5.1}) and normalized the vector
 $u^{\al}$ in eq.~(\ref{5.1}) in
 such a manner that
 \beq
\int \frac{\di^Ey}{\rho(E)} \sqrt{g^{(B)}}  u^{\al} u^{\be} g^{(B)}_{\al \be}
= 1  \label{5.14}\, .
\eeq
 By using the commutation relation
 \beq
 [\n_{\ga},\n_{\be}] u^{\alpha}  = - {R_{(B)}}^{\al}_{\delta \be
 \ga} u^{\delta}  \label{5.17} \, ,
 \eeq
 we bring the mass term~(\ref{5.16})
 into the  form:
 \bea
 m_V^2 & = & \frac{1}{2} \int \frac{\di^Ey}{\rho(E)} \sqrt{g^{(B)}}
 (\n_{\al} u_{\be} + \n_{\be}
 u_{\al} )^2  \label{5.18} \, .
\eea
Notice that to arrive at (\ref{5.18}) we used the gauge invariance
$x^{\mu} \rightarrow x^{\mu} - \epsilon^{\mu} (y)$  to fix
$\n_{\al} A_{\mu}^{\al} = 0$ (and therefore $\n \cdot u = 0$).

We thus recover the well-known result that the only way
to obtain a  massless vector by dimensional reduction of
the Einstein-Hilbert action is by having a symmetry of the
compact manifold.  This symmetry
is encoded in the existence of the Killing
vector $V^\al$ satisfying the Killing vector equation:
\beq
\n_{\al} V_{\be} + \n_{\be}
 V_{\al} = 0 \label{5.20}\, .
 \eeq
The Killing vector generates translations of the manifold $K$
along its symmetry direction. The graviphoton field describes
a translation of this kind that varies from point to point in
$D$-dimensional Minkowski-space. It is of course exactly this
local gauge invariance that brings about the masslessness of
the graviphoton.

If we restrict our attention to background field configurations
without matter energy-momentum (i.e. background manifolds
which are Ricci flat), the requirement that we have a Killing
vector is very restrictive because
\begin{itemize}

\item
 Any compact Ricci-flat manifold that admits a Killing vector
field $V^{\al}$ is flat in the direction of $V^{\al}$, i.e. satisfies
\beq
R_{\al \be \gamma \delta} V^{\al} =
R_{\al \be \gamma \delta} V^{\be} =
R_{\al \be \gamma \delta} V^{\gamma} =
R_{\al \be \gamma \delta} u^{\delta} = 0  \label{5.21}\,  ;
\eeq
moreover,
\item
The Killing vector is covariantly constant:
\beq
\n_{\al} V^{\be} = 0  \label{5.22} \, .
\eeq

\end{itemize}
The proof goes as follows.
By assumption
\beq
\int \di^Ey \sqrt{g(y)} V^{\al} V^{\be} R_{\al \be} = 0
\label{5.23}\, .
\eeq
Using the relation (\ref{5.17}) we may rewrite this in terms of a
commutator of two covariant derivatives:
\bea
0 & = & \int \di^Ey \sqrt{g} V^{\al} [\n_{\be},\n_{\al}] V^{\be}  \nnu \\
  & = & \int \di^Ey \sqrt{g} V^{\al} (\n_{\be} \n_{\al} V^{\be}
  - \n_{\al} \n_{\be} V^{\be} )  \label{5.25}
\eea
Since $V^{\al}$ is a Killing vector field,
$\n \cdot V = 0$. Next we do a partial integration to obtain
\beq
0 = \int \di^Ey \sqrt{g} \n_{\be} V^{\al} \n_{\al} V^{\be}  \label{5.26}
\eeq
Since a compact manifold has no boundary, the surface term
vanishes. Finally, using the Killing vector condition (\ref{5.20})
eq.(\ref{5.26}) is brought into the form:
\beq
0 = \int \di^Ey \sqrt{g} (\n_{\be} V^{\al})^2  \label{5.27}\, ,
\eeq
from which it follows immediately that the Killing vector is
covariantly constant.

Such a result
is very restrictive because it implies the integrability condition
\beq
[\n_{\al}, \n_{\be}] V^{\gamma} = R^{\gamma}_{\ \delta \alpha \beta}
V^{\delta} = 0  \label{5.28}\, ,
\eeq
that tells us that
 the manifold has to be flat in the direction of the Killing
vector. Note that by the symmetries of the Riemann tensor ,
(\ref{5.28}) implies
all four equations (\ref{5.21}). This concludes our proof.

If we now choose coordinates such that $V^{\al} = (1,0, \ldots,0)$ the metric
will not depend on $y^1$. By choosing Gaussian normal coordinates we can
make $g_{\al 1} = 0$ for $\al \neq 1$. Moreover,
the condition $\nabla_1
V^{\al} \equiv 0$ implies $\partial_{\al} g_{11} = 0$, i.e. by a rescaling
of the Killing vector we can obtain $g_{11} \equiv 1$.

The resulting form of the metric shows that our compact Ricci-flat manifold
$K$ of dimension $E$ decomposes into
\beq
K = S^1 \times \tilde{K} \label{5.36}
\eeq
---the direct product of a circle and a Ricci-flat manifold $\tilde{K}$
of dimension $E-1$.

The same result can be phrased by saying that
we have a Ricci-flat
manifold with SO($E-1$) holonomy.  The vector representation {\bf E}
of SO($E$) decomposes under SO($E-1$) into ({\bf E} $-$ {\bf 1})
 + {\bf 1}, and the
singlet corresponds to the existence of a covariantly constant
vector. Being covariantly constant, it is a Killing vector. As such,
it gives rise to a single massless graviphoton.

In the special case of a superstring being compactified from ten
to four dimensions,  the presence of a massless graviphoton
requires SO(5) holonomy. It is well known that to have N=1
supersymmetry in four dimensions one needs instead SU(3)
holonomy~\cite{13d}.  These two holonomy groups (both of rank two) are
clearly incompatible---neither has the other as a subgroup.
This  is hardly a surprising
result; after all, we knew that in a Calabi-Yau
compactification there are no massless graviphotons~\cite{GSW}.

\subsection{The graviscalars.}
\hspace*{2em} Having analyzed the graviphotons in  detail we now turn
our attention to the graviscalars. Relying on Ricci flatness
to remove all terms involving the background curvature, we
arrive at the following action quadratic in the graviscalar
field:
\beq
S_h^{(2)} = S^h_{kin} + S^h_{mass} \label{5.37}\, ,
\eeq
where
\beq
S^h_{kin} = -
\half \int \frac{\di^Ey}{\rho(E)} \sqrt{g^{(B)}} \di^Dx \left\{
(\n_{\mu} h^{\al \be})^2 + \frac{1}{D-2} (\n_{\mu} h)^2
\right\} \label{5.38}
\eeq
gives the kinetic part  in $D$ dimensions, and
\bea
S^h_{mass} & = &
\half \int \frac{\di^Ey}{\rho(E)} \sqrt{g^{(B)}} \di^Dx \left\{
\frac{4-D}{(D-2)^2} (\n_{\al} h)^2 + \frac{4}{D-2} \n_{\al}
h \n_{\be} h^{\al \be} \right.  \nnu \\
& & \left. + 2 \n_{\gamma} {h}^{\be}_{\al}
\n_{\be} h^{\gamma \al} - (\n_{\gamma} h^{\al \be})^2
\right\}  \label{5.39}
\eea
is the mass operator.

In order to find massless scalars, we should look for tensors $u^{\al \be}$
that are zero modes of the mass operator.
There are two obvious candidates: the background metric
\beq
u_T^{\al \be} \equiv g_{(B)}^{\al \be}  \label{5.40}\, ,
\eeq
and the tensor product of the Killing vector with itself:
\beq
u_{V}^{\al \be} \equiv V^{\al} V^{\be} \label{5.41}\, .
\eeq
Bearing in mind that the manifold has the structure $K =
S^1\times \tilde{K}$, (\ref{5.40}) and (\ref{5.41})
are easily seen to be the only zero
modes that have non-vanishing components in the direction of
the Killing vector. In addition to these we can also have some zero
modes of the compact subspace $\tilde{K}$, which we will
denote by $u^{\al \be}_i$; $(i= 1,2,\ldots)$.

Corresponding to each zero mode we have  massless graviscalar
fields that we denote by $h_T, h_V$ and $h_i$, respectively.
The only one of these graviscalars
that couples to matter charged only
under the U(1) gauge symmetry is $h_V$. Anti-gravity arises as a result of
cancellation between exchange of graviton, graviphoton and this
unique graviscalar.

To prove the decoupling of $h_T$ and $h_{i}$,
 we consider again the generic coupling to
gravity of a
 massless  field in $D+E$ dimensions, which is described by the
action (\ref{3.11}).
We take the matter field $\Phi$ to be charged under the
U(1) group of translations by the Killing vector, that is, we
fix the $y$-dependence of the field to be of the form:
\beq
\h{\Phi} (x;y) = \Phi(x) \exp \{ i p V^{\al} y_{\al} \}  \label{5.45}\, .
\eeq
We may think of $p V^{\al}$ as the
internal momentum
in the direction of the Killing vector.

As in section 3 the $D$-dimensional scalar field $\Phi$ will
have a mass
$m = |p|$ and a charge $q = \sqrt{2} \kappa p$ and will couple
to the graviton and the graviphoton in the canonical way
 given by eq.(\ref{A}) and
(\ref{B}).

The coupling to the graviscalar is given by
\beq
\kappa m^2 \int \di^D x \: \phi ^{2} (x) h(x) \, {\cal N}   \label{5.51}\, ,
\eeq
where the number ${\cal N}$ depends on the assumptions made
for the graviscalar zero mode $u_{\al \be}$:
\beq
{\cal N} = \int \frac{\di^Ey}{\rho(E)} \sqrt{g^{(B)}}
\left( u^{\al \be} V_{\al} V_{\be} + \frac{1}{D-2} g^{(B)}_{\al \be}
u^{\al \be} \right) \label{5.52}\, .
\eeq
By choosing the following basis for the zero modes
\bea
u_{V}^{\al \be} & =  & V^{\al} V^{\be} \nnu \\
u_{T}^{\al \be} & = & g_{(B)}^{\al \be} - \frac{D+E-2}{D-1} V^{\al}
V^{\be} \nnu \\
\tilde{u}_i^{\al \be} & = & u_i^{\al \be} - \frac{1}{E-1}
(g_{(B)}^{\al \be} - V^{\al} V^{\be}) (g^{(B)}_{\gamma \delta}
u_i^{\gamma \delta}) \label{5.53}\, ,
\eea
we find that all scalar
states decouple (i.e.  ${\cal N} = 0$), except
the one  related to the mode $u_V$.  Furthermore, the graviscalar
propagator given by eq.(\ref{A.3}) is diagonalized by the
choice of basis~(\ref{5.53}) . The propagator of this
graviscalar is given by:
\beq
 \langle h_V(p) h_V (-p) \rangle =
  \frac{1}{p^2} \frac{D-2}{D-1}  \label{5.54}\, .
\eeq
Hence, we are back to the case of toroidal compactification even for the
graviscalars and the computation of the four-point amplitude proceeds as
in section 3, with the resulting  eq.(\ref{3.16}).

The solution  of Einstein's equations will not be Ricci flat
if we introduce
 a classical matter distribution on the compact manifold.
It is then
 possible to have Killing vectors which are not covariantly
constant and form non-Abelian groups. The resulting massless vectors in
$D$ dimensions will
have couplings that depend on the details of the manifold,
i.e. on the matter distribution introduced~\cite{WC}. There will
in general be  no
anti-gravity. This is particularly clear if we consider the graviscalar
$h_V$,
because the mode (\ref{5.41}) from which it is constructed is not a zero
mode of the mass operator (\ref{5.39})  unless $V^{\al}$ is covariantly
constant.

We conclude that  the anti-gravity phenomenon is a special property of
toroidal compactification.

\section{Conclusions.}

\hspace*{2em} We have seen that anti-gravity is
 a feature of any four-dimensional effective
theory of gravity obtained from a higher dimensional
theory by compactification on a manifold with flat directions.

Anti-gravity occurs between states that
start out by being  massless in the higher-dimensional
theory but obtain on compactification a mass and
a charge under the U(1) group of rigid translations in
the flat direction.

If these states obtain their U(1) charge by virtue of
a non-vanishing compact momentum, they
 are going to be  very heavy (of the order of, say, $ 10^{16}$
GeV) and hence of little relevance to experimental physics (except,
perhaps, as dark-matter candidates).
However, as it has been
shown by Scherk~\cite{Scherk}, U(1) charged states need not be very heavy
and can appear in the process of supersymmetry breaking.  This leaves
open the most important question about anti-gravity, that is,
whether
ordinary matter is charged or not under the internal U(1) group
of the graviphoton. If ordinary matter turns out to be charged,
the mass of the graviscalar and the graviphoton cannot be truly zero,
since this would violate many experimental
bounds on the principle of equivalence~\cite{Will}.  Among
fifth-force candidates,  anti-gravity is singled out by
the prediction that the new force couples with the same strength of
gravity.  While a more complete discussion of the experimental relevance
of anti-gravity is outside the scope of this paper, we want to stress
here that, as we have seen in the previous sections,
anti-gravity is the low-energy signature of having a flat compact
dimension.  Flat compact directions are incompatible with
 Calabi-Yau compactifications and will tend to produce
too much supersymmetry in four dimensions, making it very hard to
accommodate the chiral asymmetry of the real world.

\section{Acknowledgement.}
\hspace*{2em} Many thanks to Pietro Fr\'{e} and Roberto Iengo for
useful discussions.

\appendix
\section{Field equations of compactified Einstein theory.}
\setcounter{equation}{0}
\hspace*{2em}
In this appendix we present the full field equations of the theory
(\ref{3.6}) in the presence of point-like sources.

First, there are the Einstein equations for the gravitational field:
\beq
R_{\mu \nu} - \half g_{\mu \nu} R  =  \kappa^2 T_{\mu \nu} \, , \label{B.1} \\
\eeq
where the energy-momentum tensor splits into two parts. The
first part is due to the energy-momentum of the graviscalar and graviphoton
fields.
\bea
T_{\mu \nu}^{(\phi,A)} & = & \delta^{-\gamma} \left( F^{\be}_{\mu \rho}
F_{\nu}^{\ \rho \al} \phi_{\al \be} - \frac{1}{4} g_{\mu \nu} F^{\rho \tau
\beta} F_{\rho \tau}^{\al} \phi_{\al \be} \right) \nnu \\
& & + \frac{1}{4 \kappa^2} \phi^{\al_1 \al_2} \phi^{\be_1 \be_2}
\left( \partial_{\mu} \phi_{\al_1 \be_1} \partial_{\nu} \phi_{\alpha_2
\beta_2} - \half g_{\mu \nu} \partial_{\rho} \phi_{\al_1 \be_1} \partial^{\rho}
\phi_{\al_2 \be_2} \right) \nnu \\
& & +\frac{1}{4\kappa^2 (D-2)} \left( \partial_{\mu} \log \delta
\: \partial_{\nu}
\log \delta - \half g_{\mu \nu} (\partial_{\rho} \log \delta)^2 \right)
\label{B.2} \, .
\eea
The second part is due to the motion of a point-particle source of canonical
momentum $\h{\pi}_{\mu}$ and charge $q_{\al}$:
\bea
T_{\mu \nu}^{(PP)} & = & \delta^{-\gamma} \int \di \sigma (\h{\pi}_{\mu}
+ q_{\be} A^{\be}_{\mu}) (\h{\pi}_{\nu}
+ q_{\al} A^{\al}_{\nu}) \frac{\delta^D(x - x(\sigma))}{\sqrt{-g}}
\label{B.3}\, .
\eea
The trajectory $x(\sigma)$
of the particle is related to the canonical momentum by
\beq
g_{\mu \nu} \frac{\di x^{\nu}}{\di \sigma} = \delta^{-\gamma} (\h{\pi}_{\mu}
+q_{\be} A_{\mu}^{\be} ) \label{B.4a}\, .
\eeq
Here $\sigma = \tau/m_D$, $\tau$ being the proper time and $m_D$ the rest
mass of the point particle.

The equations for the graviphoton field are
\beq
\n_{\rho} (\delta^{-\gamma} F^{\rho \mu \beta}\phi_{\al \beta} ) =
\delta^{-\gamma}
J^{\mu}_{\al}  \label{B.4}\, ,
\eeq
where the charge current density of the source is given by
\beq
J^{\mu}_{\al} = \int \di \sigma q_{\al} (\h{\pi}^{\mu} + q_{\be} A^{\mu \beta}
) \frac{\delta^D(x-x(\sigma))}{\sqrt{-g}} \label{B.5}\, .
\eeq
Finally, we have the graviscalar equations of motion
\bea
& \Box \phi_{\al \be} + \frac{1}{D-2}
\phi_{\al \be} \Box \log \delta
-\phi^{\al_1 \al_2} \n_{\rho} \phi_{\be \al_2} \n^{\rho} \phi_{\al \al_1}
 & \nnu
\\
&  + \kappa^2 \delta^{-\gamma} \left(
 F^{\al _1}_{\mu \nu}
F^{\beta_1 \mu \nu} \phi_{\al \be} \phi_{\al _1 \be _1}
 - F^{\mu \nu \al _1} F_{\mu \nu}^{\beta_1} \phi_{\al _1 \al}
 \phi_{\beta_1 \be}
\right)  = & \nnu \\
&  +\frac{2\kappa^2}{D-2}  \phi_{\al \be} T^{(PP)} - \Pi_{\al \be} \, , &
\label{B.6}
\eea
where $\Box$ is the curved-space Laplacian and the charge matrix density
is given by
\beq
\Pi_{\al \be} = \int \di \sigma q_{\al} q_{\be}
\frac{\delta^D(x-x(\sigma))}{\sqrt{-g}} \label{B.7}\, .
\eeq
In the above equations $\delta$ and $\gamma$ are given by eqs.(\ref{3.4})
and (\ref{3.5}), respectively.

The solution (\ref{4.11}) corresponds to a point-particle source
of mass $M$ and
charge
\beq
q_{\al} = \epsilon \sqrt{2} \kappa M \delta_{\al}^1   \label{B.8}
\eeq
stationary at the origin.
It is a solution of the full field equations of $N=8$
(or $N=4$) supergravity because,
of the many fields appearing in that theory, only those contained in
(\ref{3.6}) are excited by the source (\ref{B.8}).

The equations of motion of a test particle of charge
$q_{\be}$ and mass $m_D$ in the above theory are derived
within the $D$-dimensional framework by requiring conservation of
energy-momentum of field + test-particle:
\beq
\n_{\nu} (T^{\mu \nu}_{(\phi,A)} + T^{\mu \nu}_{(PP)} ) = 0 \, .\label{B.9}
\eeq
The equations are the following ones:
\bea
\lefteqn{ \stackrel{..}{x}^{\mu} + \Gamma^{\mu}_{\rho \tau} \dot{x}^{\rho}
\dot{x}^{\tau} =  - \frac{q_{\be}}{m_D} \delta^{-\gamma}
F^{\mu \be}_{\ \: \rho} \dot{x}^{\rho} }
\nnu \\
& & + \frac{1}{2(D-2)} \phi^{\al \be} ( \delta^{\mu}_{\tau} \partial_{\rho}
+ \delta^{\mu}_{\rho} \partial_{\tau} - g_{\rho \tau} \partial^{\mu} )
\phi_{\al \be} \dot{x}^{\rho} \dot{x}^{\tau} - \frac{1}{4 \kappa^2 m_D^2}
q_{\al} q_{\be} \partial^{\mu} \phi^{\al \be} \label{B.10}\, ,
\eea
where the dot denotes differentiation with respect to $\tau$.

\newpage
\renewcommand{\baselinestretch}{1}

\end{document}